\documentclass[twocolumn,prx,aps,10pt]{revtex4-1}

\usepackage{amsfonts,amssymb}
\usepackage[sumlimits,intlimits]{amsmath}
\usepackage{graphics}
\usepackage{graphicx}
\usepackage{mathrsfs}
\usepackage{textcomp}
\usepackage{verbatim}
\usepackage[colorlinks = true,		
            linkcolor = blue,
            urlcolor  = blue,
            citecolor = blue,
            anchorcolor = blue,
            bookmarks=false]{hyperref}

\usepackage{bm}          

\newcommand{\be}{\begin{equation}}
\newcommand{\ee}{\end{equation}}
\newcommand{\copt}{C_\text{opt}}
\newcommand{\ceq}{C_\text{eq}}

\newcommand{\xipl}{\xi_\parallel}
\newcommand{\xipd}{\xi_\perp}
\newcommand{\nupl}{\nu_\parallel}
\newcommand{\nupd}{\nu_\perp}

\begin{document}

\title{The price of anarchy is maximized at the percolation threshold}

\author{Brian Skinner}
\affiliation{Materials Science Division, Argonne National Laboratory, Argonne, IL 60439, USA}

\date{\today}

\begin{abstract}

When many independent users try to route traffic through a network, the flow can easily become suboptimal as a consequence of congestion of the most efficient paths.  The degree of this suboptimality is quantified by the so-called ``price of anarchy" (POA), but so far there are no general rules for when to expect a large POA in a random network.  Here I address this question by introducing a simple model of flow through a network with randomly-placed ``congestible" and ``incongestible" links.  I show that the POA is maximized precisely when the fraction of congestible links matches the percolation threshold of the lattice.  Both the POA and the total cost demonstrate critical scaling near the percolation threshold.

\end{abstract}

\maketitle

\section{Introduction}

Optimizing the flow through a large network is no easy task.  This fact is readily apparent to anyone who has lost their internet connection or been stuck in a traffic jam.  The problem in such cases, and in a myriad of others, is that network links become congested when a large demand is placed on them, and this congestion lowers the network efficiency for all users.  It is thus common for situations to arise where the self-interest of individual users, who would prefer to route their traffic through the most convenient links, is not aligned with the globally optimum behavior for all users, which generally requires that traffic be distributed more evenly to avoid congestion.  Understanding this self-interest-driven inefficiency is critical for a large number of problems across an array of fields, including the planning of computer networks\cite{roughgarden_selfish_2005}, transportation networks\cite{braess_uber_1968, wardrop_road_1952, youn_price_2008, zhu_measuring_2010}, and power grids\cite{witthaut_braesss_2012}, the allocation of public services\cite{knight_selfish_2013}, and even sports strategy\cite{skinner_price_2010, skinner_price_2013}.  

Unfortunately, at present there are no general rules for predicting which conditions give rise to large inefficiency in a random network.  Indeed, previous works have largely studied congestible networks either at the level of proving general theorems\cite{valiant_braesss_2010, roughgarden_how_2002, roughgarden_price_2003} or by trying to model specific real-world systems as accurately as possible\cite{tsekeris_design_2009, youn_price_2008}.  In this paper, I seek a middle ground approach by introducing and analyzing a simple lattice model of randomly-placed congestible and incongestible links.  The model takes inspiration from the spectacular history of similarly idealized lattice models in theoretical physics, which have provided important and generic insights for fields as diverse as electrical conduction in disordered materials\cite{fisch_critical_1978, skal_topology_1975, nguen_tunnel_1985}, the spread of epidemics\cite{moore_epidemics_2000, sander_percolation_2002}, the mechanics of biopolymer networks\cite{broedersz_criticality_2011}, and the quantum Hall effect\cite{chalker_percolation_1988}.  Crucially, in each of these examples the lattice percolation threshold plays a key role in determining the behavior of the system.  As I show below, the same is true for the model proposed here.  

For the sake of concreteness, the remainder of this paper discusses the network using the language of cars on a road network\footnote{such language may become particularly relevant with the further development of self-driving cars\cite{markoff_google_2010}.}, but one can think of the words ``roads," ``traffic," and ``commute time," as equivalent to the more generic concepts of ``links," ``current," and ``cost."

\section{Proposed Model}
\label{sec:model}

\begin{figure}[htb]
\centering
\includegraphics[width=.85\columnwidth]{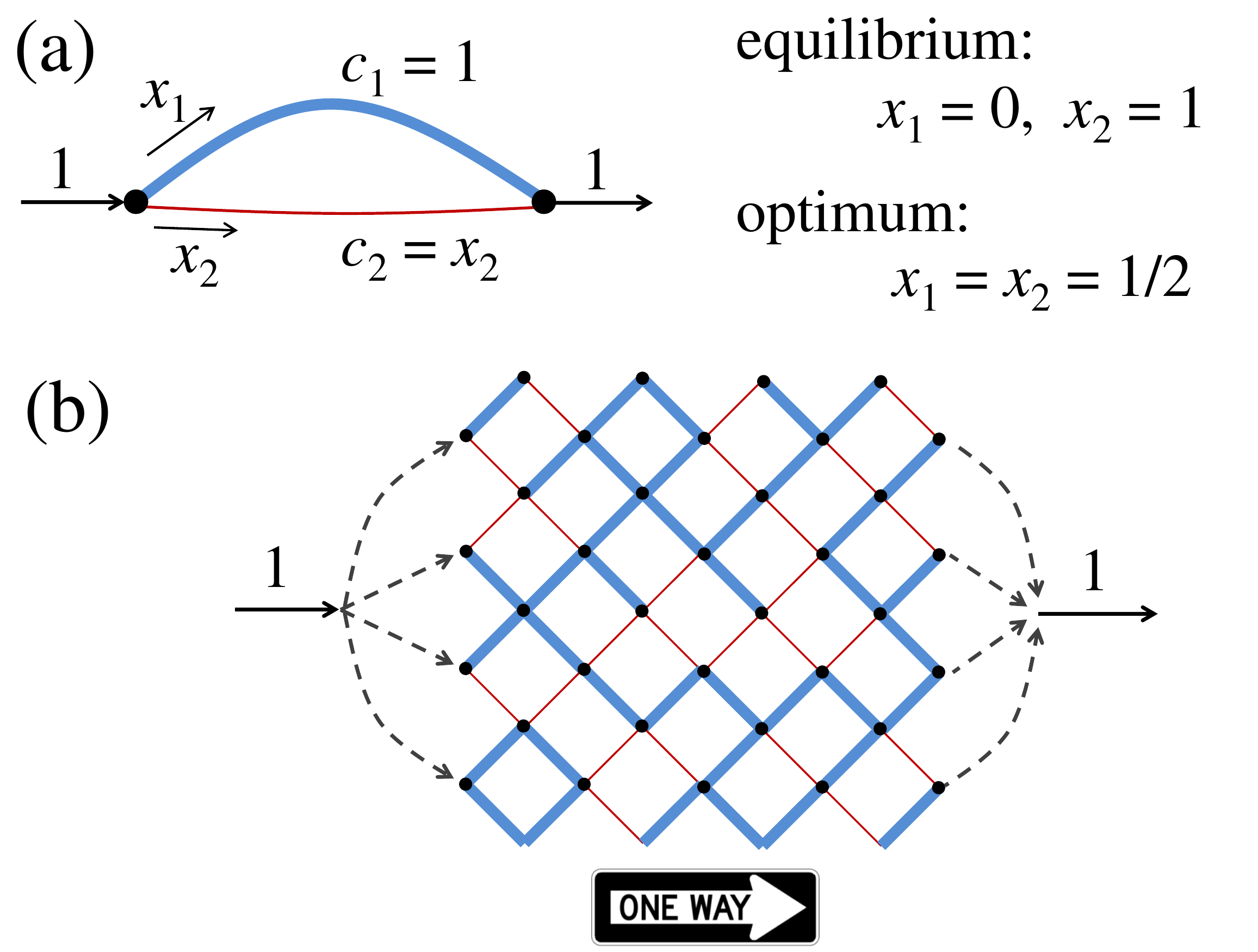}
\caption{(Color online) (a) Pigou's example of flow through a network with two ``roads," one of which has constant commute time and the other of which is congestible.   (b) The model introduced here, where constant roads (thin red lines) and congestible roads (thick blue lines) are randomly mixed.  A unit amount of traffic is passed through the network, and all roads are unidirectional and proceed to the right.}
\label{fig:schematics}
\end{figure}

The canonical example of network suboptimality arising from user self-interest was introduced by the economist Arthur Pigou\cite{pigou_economics_1924}, who in 1920 imagined the simple network of two parallel roads shown in Fig.\ \ref{fig:schematics}(a).  In Pigou's example, one of the two roads offers a constant commute time $c_1 = 1$ (in some units).  The second road offers a potentially much faster commute but is highly congestible.  This road is described by a usage-dependent commute time $c_2 = x_2$ (in the same units), where $x_2 \in [0, 1]$ denotes the proportion of the total traffic on this second road.  It is straightforward to show that the average commute time per driver, $C = x_1 c_1 + x_2 c_2(x_2) = x_1 + x_2^2$, where $x_1 = 1 - x_2$ is the proportion of the traffic taking the first road, is optimized when the two roads are used equally: $x_1 = x_2 = 1/2$.  This optimum arrangement has an average commute time $\copt = 3/4$.

The optimum configuration, however, is unstable in a social sense: any driver on the first road experiences a commute time of $c_1 = 1$, while drivers on the second road experience a twice-shorter commute $c_2 = 1/2$.  Thus, in the absence of any regulation, all drivers on the first road have an incentive to switch to the second road, and the resulting ``equilibrium" has $x_1 = 0$, $x_2 = 1$, and $C = \ceq = 1$.  This equilibrium is equivalent to the Nash equilibrium from game theory, which is generally defined as the state in which no user can improve his commute time by changing his commute path \cite{wardrop_road_1952}.

The inefficiency of the equilibrium is quantified by the so-called ``price of anarchy" (POA), $P = \ceq/\copt \geq 1$; one can think that the POA reflects the increase in global cost that results from allowing users to choose their own paths rather than having the optimum arrangement dictated to them.  

While Pigou's example provides an illustration of self-interest-driven inefficiency, there is no general rule for when to expect a significant POA in a large network.  Motivated by this lack of qualitative understanding, I introduce a generalization of Pigou's example in which slow ``constant roads," having $c(x) = 1$, are combined with faster ``congestible roads," having $c(x) = x$, to form a large network.  The majority of this paper focuses on the case where roads are arranged into a two-dimensional square lattice, as shown in Fig.\ \ref{fig:schematics}(b).  All roads in the lattice are taken to be uni-directional, and a unit amount of traffic is passed through the network, such that all traffic enters through the $L$ vertices on the left boundary and exits through the $L$ vertices on the right boundary.  The lattice is taken to be periodic in the vertical direction, and in this configuration each of the $2^{2L}$ possible paths through the network involves traversing exactly $2L$ roads.  Each road type is assigned randomly, with a probability $p$ of being of the congestible type.

The main focus of this paper is on calculating the global average commute time,
\be 
C = \sum_{\text{all roads }i} x_i c_i(x_i),
\label{eq:C}
\ee 
for both the optimum and equilibrium configurations as a function of $p$ and the system size $L$.  As outlined above, the traffic $x_i$ on a given road $i$ satisfies $0 \leq x_i \leq 1$ and the commute time across the road is described by either $c_i(x_i) = 1$ or $c_i(x_i) = x_i$. 

A nontrivial dependence of the POA as a function of $p$ can be anticipated by looking at the extreme cases of $p = 0$ and $p = 1$.  At $p = 0$, all roads have constant cost $c = 1$ regardless of the arrangement of traffic, and so $\ceq = \copt = 2L$ and $P(0) = 1$.  On the other hand, at $p = 1$ all roads have $c = x$.  In this case the symmetry of the lattice demands that all roads have equal usage $x = 1/2L$ in both the equilibrium and optimum conditions, and as a consequence $\ceq = \copt = 1$ and $P(1) = 1$.  For the case $0 < p < 1$ however, where the network is non-uniform, one can expect $P - 1$ to be finite, as fast but congestible roads get overused in the equilibrium.

\section{Numerical Modeling and Electrical Circuit Analogy}
\label{sec:numerics}

In order to find the total average commute time $C$ in either the equilibrium or optimum situation, one should first solve for the traffic $x_i$ along each road $i$.  For the optimum case, this solution amounts to finding the minimum of the quadratic function $C$ with respect to all variables $x_i$ [see Eq.\ (\ref{eq:C})], subject to the constraints that $x_i \geq 0$ for all $i$, that the current is conserved at each node, and that the total current passed through the network is equal to unity.  Similarly, the equilibrium can be thought of as the minimum of the modified cost function $\sum_i \int_0^{x_i} c_i(x') dx'$, subject to the same constraints.  This equilibrium is equivalent to the state where all possible commute paths across the network have either the same commute time or have zero usage.  Since finding the equilibrium or optimum therefore represents a minimization of a convex second-degree polynomial over a convex domain, the solution can be found using standard algorithms from convex optimization.\cite{dasgupta_algorithms_2008}  Details about the algorithm used here are given below.

\begin{figure}[htb!]
\centering
\includegraphics[width=\columnwidth]{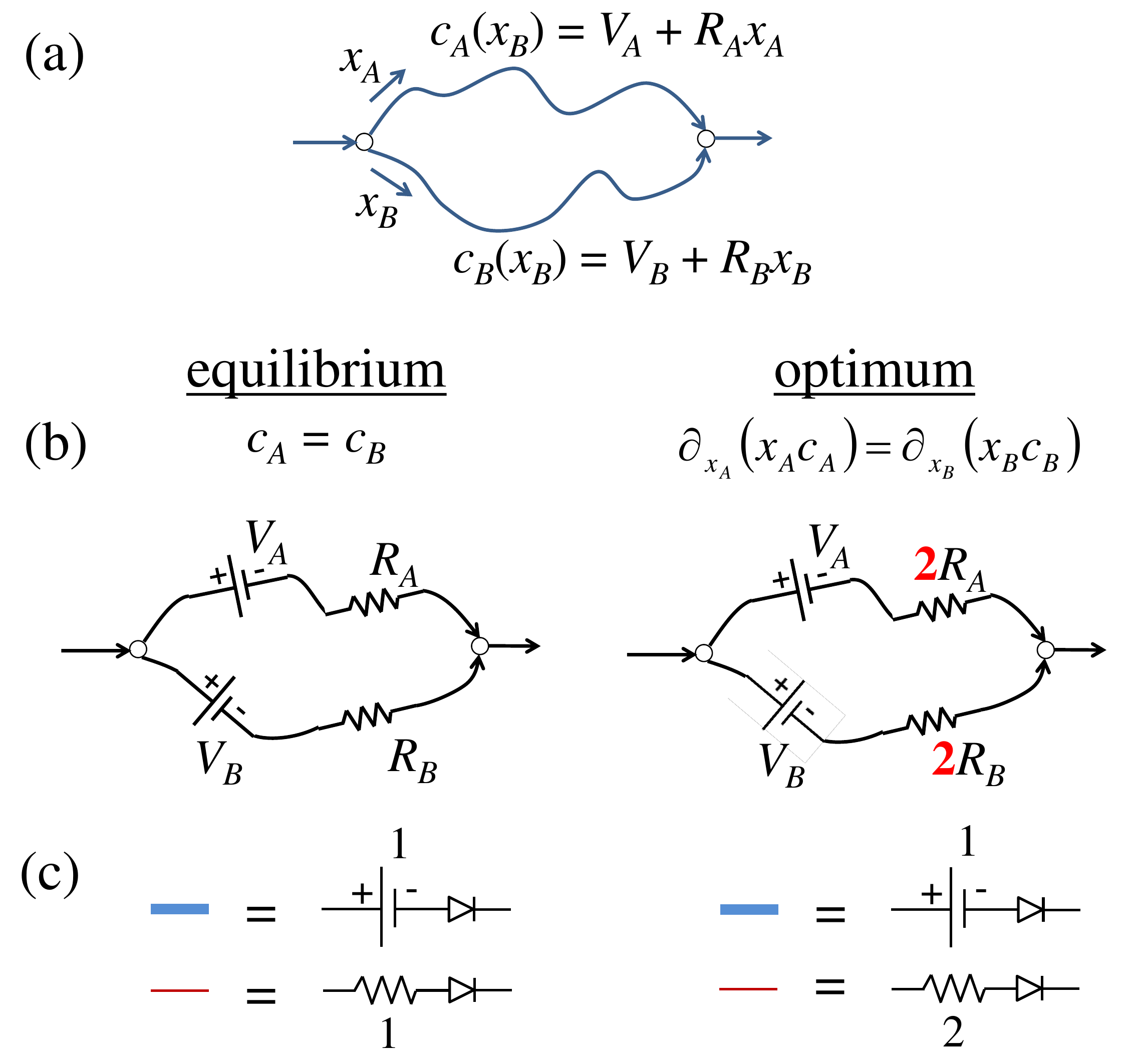}
\caption{(Color online) Electrical circuit analog of the lattice shown in Fig.\ \ref{fig:schematics}(b).  (a) Traffic splits between two paths, which generically have commute times $c_A$ and $c_B$, respectively, that are linear functions of the traffic along each road.  (b) The equilbrium traffic, for which $c_A = c_B$, is analogous to the current through a resistor network with resistances matching the value of $\partial c/\partial x$ along each branch.  The optimum traffic, on the other hand, is analogous to the current resulting from a circuit with \emph{doubled} values of the resistance.  Currents are assumed to proceed to the right.  (c) The network model being considered [as in Fig.\ \ref{fig:schematics}(b)] can generically be mapped onto an electrical circuit by representing constant roads as unit voltage sources and congestible roads as either unit resistors, for the equilibrium case, or as resistors with resistance $2$, for the optimum case.  The diodes ensure that current proceeds to the right, and that all commute times are positive.}
\label{fig:circuit-schematic}
\end{figure}

Some insight into the problem of solving for the network flow can be gained by recasting it in the more familiar language of a linear electrical circuit.  In this analogy, the traffic $x$ along a given road is equated with a ``current", and the commute time $c(x)$ is equated with a ``voltage drop".  Thus, the constant roads with $c(x) = 1$ are equivalent to links with unit voltage sources, which produce a current-independent voltage drop across the link.  The congestible roads with $c(x) = x$ are analogous to links with unit resistors, which produce a voltage drop that increases linearly with the current.  Importantly, each link also includes an (ideal) diode [as depicted in Fig.\ \ref{fig:circuit-schematic}(c)], which enforces the unidirectionality of the current.  Such diodes ensure that the voltage drop across each link is strictly positive, or, equivalently, that all roads require a positive amount of time to traverse.

Once this analogy is made, the current on each link can be solved for using the corresponding Kirchoff laws.  In particular, the Kirchoff loop law states that for any two paths $A$ and $B$ connecting the same two points in the network, the total voltage drops $c_A$ and $c_B$ along the paths are equal.  This is equivalent to the Nash equilibrium condition, which states that any two paths with finite usage must have the same commute time; otherwise commuters will switch from the slower path to the faster one.  

On the other hand, the optimum currents $x_A$ and $x_B$ along the two paths are not those which equalize $c_A$ and $c_B$, but are instead those which minimize the average commute time, $x_A c_A + x_B c_B$, subject to the constraint that $x_A + x_B$ is fixed.  For nontrivial solutions having $x_A, x_B > 0$, this corresponds to the condition $\partial_{x_A}(x_A c_A) = \partial_{x_B}(x_B c_B)$.   Consider, for example, that the commute time along path $A$ is described by the linear function $c_A(x_A) = V_A + R_A x_A$, where $V_A$ and $R_A$ are coefficients that represent the number of constant or congestible roads, respectively, along path $A$.  Let the commute time along path $B$ be similarly written $c_B(x_B) = V_B + R_B x_B$.  In this case it is easy to show that a nontrivial optimum satisfies $V_A + 2 R_A x_A = V_B + 2 R_B x_B$.  But this relation is precisely the Kirchoff loop law for the case where the ``resistances" $R_{A/B}$ are doubled.  Thus, one can solve for the optimum currents simply by modeling each congestible road as a resistor with resistance $2$ rather than as a unit resistor, as depicted in Fig.\ \ref{fig:circuit-schematic}(b).  [Note, for example, that such a substitution gives the optimum currents $x_A = x_B = 1/2$ in Pigou's example, Fig.\ \ref{fig:schematics}(a).] This result, while elementary, already has important implications for network routing: it implies that traffic through a linear, congestible network is optimized when users respond to a cost that rises twice as fast with increased usage as does the bare transit time cost.  The possibility of achieving such a cost rate doubling in real networks is already being studied; for example, on major roadways it might be achievable through through dynamic tolling.\cite{tsekeris_design_2009}

Of course, for the problem considered here, the process of solving for the circuit currents is complicated by the presence of diodes on each link (or, equivalently, by the constraint that all commute times be strictly positive).  This condition necessitates the use of a more careful procedure than a simple solution of Kirchoff's equations, since solving for the current in all circuit elements requires one to simultaneously ascertain the correct state of all $(2L)^2$ diodes.  To find this solution, I use here a simple adaptation of the ``greedy" algorithm that is commonly used to search for the ground state of spin or Coulomb glasses\cite{shklovskii_electronic_1984}.  

In particular, in this algorithm the initial state of each diode (``on" or ``off") is first guessed randomly, and the corresponding system of Kirchoff equations is solved numerically.  The resulting currents and voltages coming from this initial solution have, in general, a number of violations of the assumed states of the diodes.  That is, some diodes assumed to be ``on" have negative currents in the (erroneous) solution, and some diodes assumed to be ``off" have positive voltage drops across them.  In the greedy algorithm, the ``on" diode with the largest negative current and the ``off" diode with the largest positive voltage both have their states switched, and the Kirchoff laws are re-solved.  This process is continued until a solution is reached that has complete consistency between the currents and the diode states.  A uniqueness theorem for linear circuits with diodes\cite{black_resistive_1960} guarantees that any such solution is unique in terms of its dissipated power, which is the analogue of the total commute time $C$.  If no such solution is reached after a large number of numerical iterations (which happens rarely), then the process is re-initialized using a different initial guess for the diode states.

This procedure is implemented numerically for each random realization of the network in order to calculate values of the traffic $\{x_i\}$ on all of the system's roads in both the optimum and equilibrium situations.  The resulting total commute times, $\copt$ and $\ceq$, are then calculated using Eq.\ (\ref{eq:C}).  Results presented below correspond to averages over many random networks for each value of $p$ and each system size $L$.

\section{Results}
\label{sec:results}

\begin{figure}[htb]
\centering
\includegraphics[width=\columnwidth]{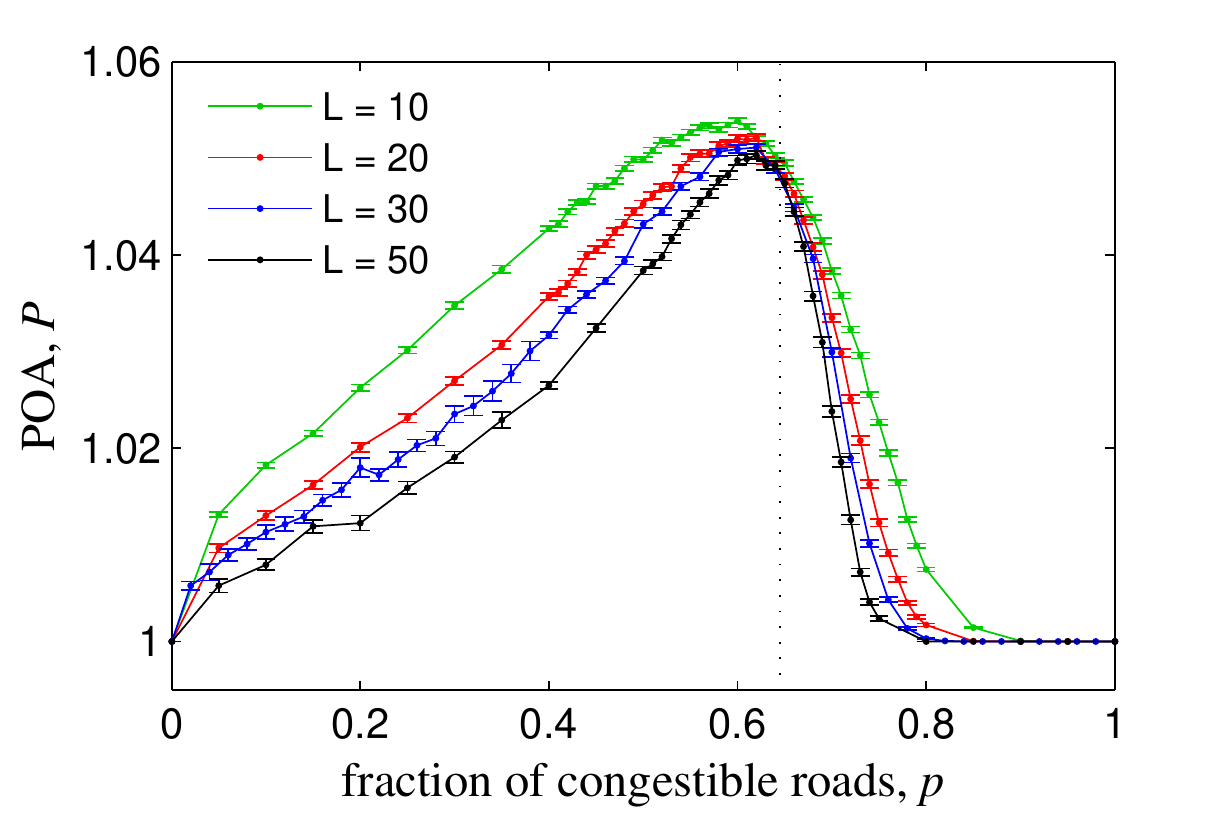}
\caption{(Color online) POA as a function of the fraction $p$ of congestible roads in the lattice, plotted for different values of the system size $L$. The vertical dotted line indicates the lattice percolation threshold, $p_c \approx 0.6447$.}
\label{fig:POA}
\end{figure}

The results for the POA are shown in Fig.\ \ref{fig:POA} as a function of $p$ for a range of system sizes.  Notably, the POA acquires its maximum value very close to the point where $p$ matches the percolation threshold for directed percolation through the lattice, $p_c \approx 0.6447$.\cite{jensen_low-density_1999}  As the system size is increased, the maximum of $P$ becomes increasingly sharp and moves closer to $p = p_c$.  As shown below, scaling of the curves $P(p, L)$ suggests that in the limit of infinite system size, the POA peak is infinitely sharp and is located precisely at $p = p_c$.

While the focus of this paper is on the square lattice shown in Fig.\ \ref{fig:schematics}(c), I briefly note that one can verify the generality of the main result of this paper, that the POA is maximized at the percolation threshold, by examining other lattice types with different values of $p_c$.  Such lattices can be expected to produce qualitatively similar curves $P(p)$ as in Fig.\ \ref{fig:POA}, but with the maximum shifted to the percolation threshold of the lattice being considered.  This is demonstrated explicitly in Appendix \ref{app:3D} for one specific case.

\begin{figure}[htb]
\centering
\includegraphics[width=.95\columnwidth]{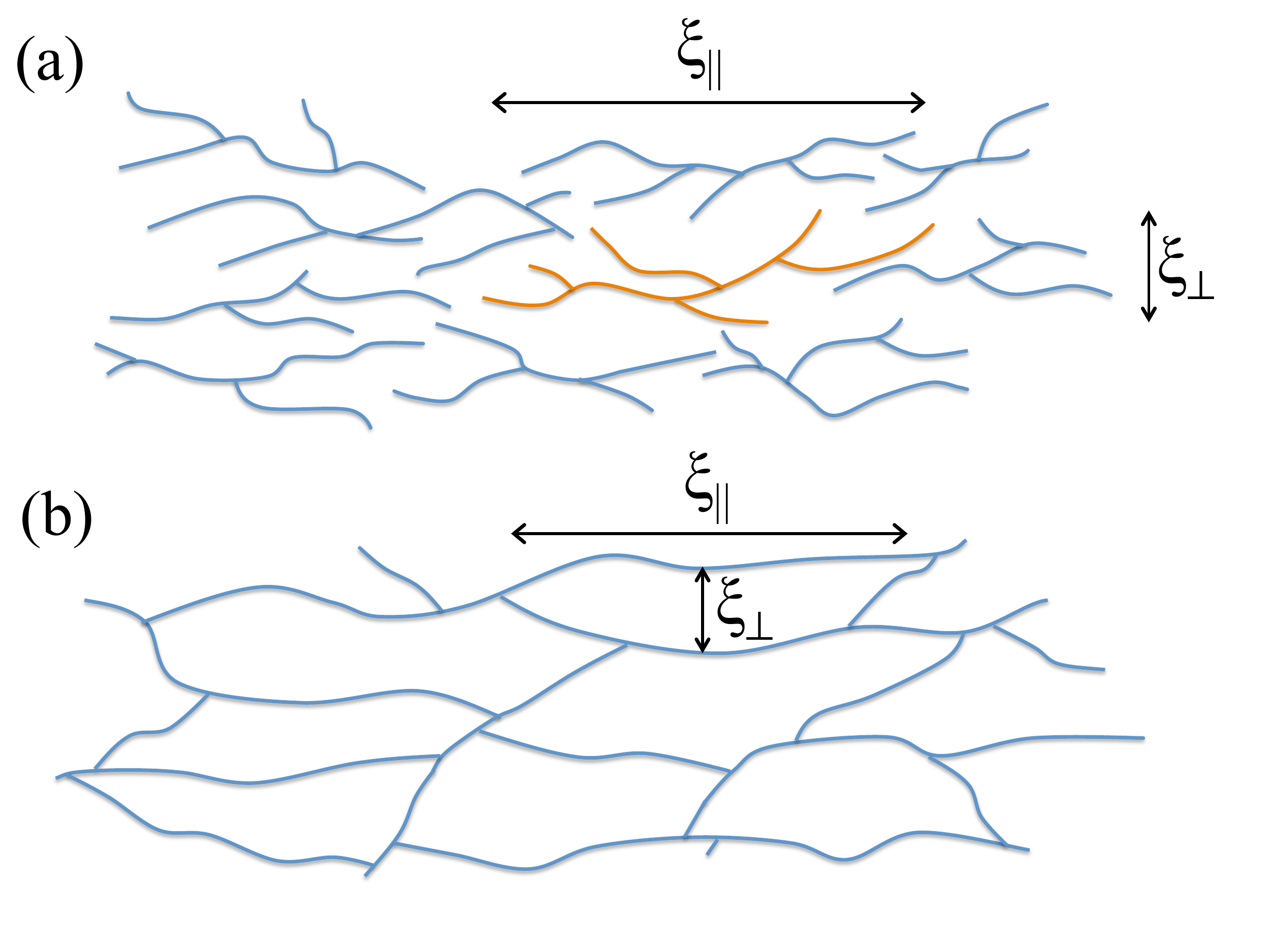}
\caption{(Color online) Schematic illustration of large clusters and percolating pathways of congestible roads.  (a) At $p < p_c$ and $|p - p_c| \ll 1$, the system contains large, disconnected clusters of congestible roads, with typical size $\xipl$ in the downstream direction and $\xipd$ in the perpendicular direction.  One such cluster is highlighted in orange. (b) At $p$ slightly larger than $p_c$, on the other hand, there are many parallel pathways for traversing the lattice using only congestible roads.  The correlation lengths $\xipl$ and $\xipd$ describe the typical horizontal and vertical separation between these paths.  Small, isolated clusters and ``dead ends" are not shown.}
\label{fig:clusters}
\end{figure}

That the POA achieves a maximum at the percolation threshold can be rationalized using the following qualitative argument.  At $p < p_c$, there are no continuous paths in a large system that connect opposite faces of the system while traversing only congestible roads, as illustrated in Fig.\ \ref{fig:clusters}(a).   Thus, all traffic across the network must use a combination of congestible roads and the slower constant roads.  In this situation the equilibrium traffic is naturally distributed over many paths with relatively low susceptibility to congestion, and the POA is not too large.  Exactly at the percolation threshold, $p = p_c$, there appears a single macroscopic pathway (the ``infinite cluster," in percolation language\cite{shklovskii_electronic_1984}) connecting opposite sides of the system that uses only congestible roads.  This single pathway becomes heavily over-used in the equilibrium, and the POA is relatively large.

\begin{figure}[htb]
\centering
\includegraphics[width=.75\columnwidth]{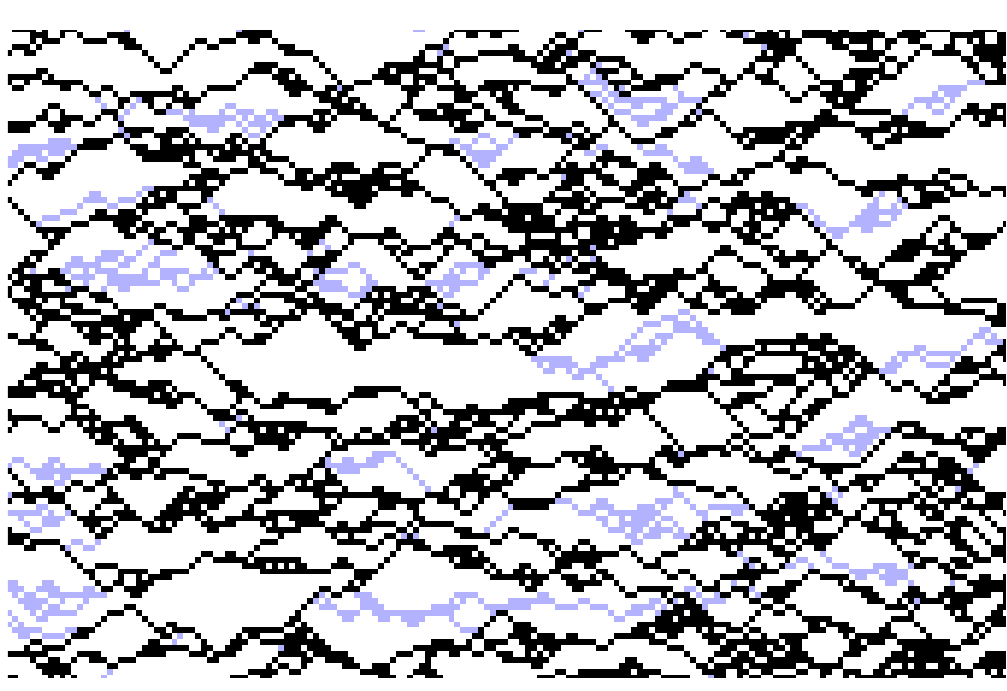}
\caption{(Color online) Spatial map of the traffic flow in a typical realization of a random network at $p \approx p_c$. Roads that carry finite traffic in the equilibrium are shown as black points, and light blue points indicate additional roads that are used in the optimum. The image has been cropped vertically, but shows the full width ($L = 75$) of the system.}
\label{fig:currents}
\end{figure}

Finally, when $p > p_c$ there are many pathways connecting opposite faces that use only congestible roads, as shown in Fig.\ \ref{fig:clusters}(b), and as $p$ is increased the number of such pathways increases.  At such large $p$ the constant roads are abandoned in the equilibrium, and the POA is a reflection only of the degree of congestion on those pathways that have finite equilibrium usage.  Increasing $p$ eases that congestion, and the POA falls abruptly.  Fig.\ \ref{fig:currents} shows a visualization of the traffic through the network at $p \approx p_c$ in both the equilibrium and optimum configurations.  As one can see, the traffic is distributed over more paths in the optimum than in the equilibrium.  Additional visualizations of the traffic density are given in Appendix \ref{app:maps}.

In order to understand the scaling behavior of the equilibrium and optimum commute times, let us first consider the case where the system size $L \rightarrow \infty$, while $p - p_c$ remains finite.  At $p < p_c$ and $|p - p_c| \ll 1$, the network contains large but disconnected clusters of fast, congestible roads with maximum size $\xipl \propto |p - p_c|^{-\nupl}$ in the downstream direction and $\xipd \propto |p - p_c|^{-\nupd}$ in the perpendicular direction\cite{family_relation_1982, redner_directed_1982}, as illustrated in Fig.\ \ref{fig:clusters}(a).  Here, $\nupl \approx 1.733$ and $\nupd \approx 1.097$ are critical exponents\cite{jensen_low-density_1999}, with $\nupl > \nupd$ indicating larger correlation length in the downstream direction.  (This behavior can also be seen qualitatively in Fig.\ \ref{fig:currents}, which shows that near the percolation threshold large, asymmetric holes appear in the current paths.)

As traffic passes from the left to the right side of the network, it generally seeks to avoid the slow constant roads, and therefore it preferrentially follows the ``backbone" of large percolation clusters.  However, since these clusters are not connected macroscopically, the traffic must pass through at least one constant road each time it moves from one cluster to another.  Therefore, the traffic generally passes through $\sim 1$ such road for each path length $\xipl$ travelled.  Since only a unit amount of current is passed through the entire system, the amount of traffic through any given percolation cluster is vanishingly small in the limit $L \rightarrow \infty$, and the commute time across the congestible roads within the cluster is also vanishingly small.  Consequently, the total commute time $C$ is dominated by passage through the constant roads connecting adjacent clusters.  Since commuting across the entire lattice requires drivers to pass from one large cluster to another $\sim L/\xipl$ times, one can expect $C$ to scale as
\be 
C \sim \frac{L}{\xipl} \sim L|p-p_c|^{\nupl}
\label{eq:Cpsmall}
\ee 
at $p < p_c$.

\begin{figure}[htb]
\centering
\includegraphics[width=.95\columnwidth]{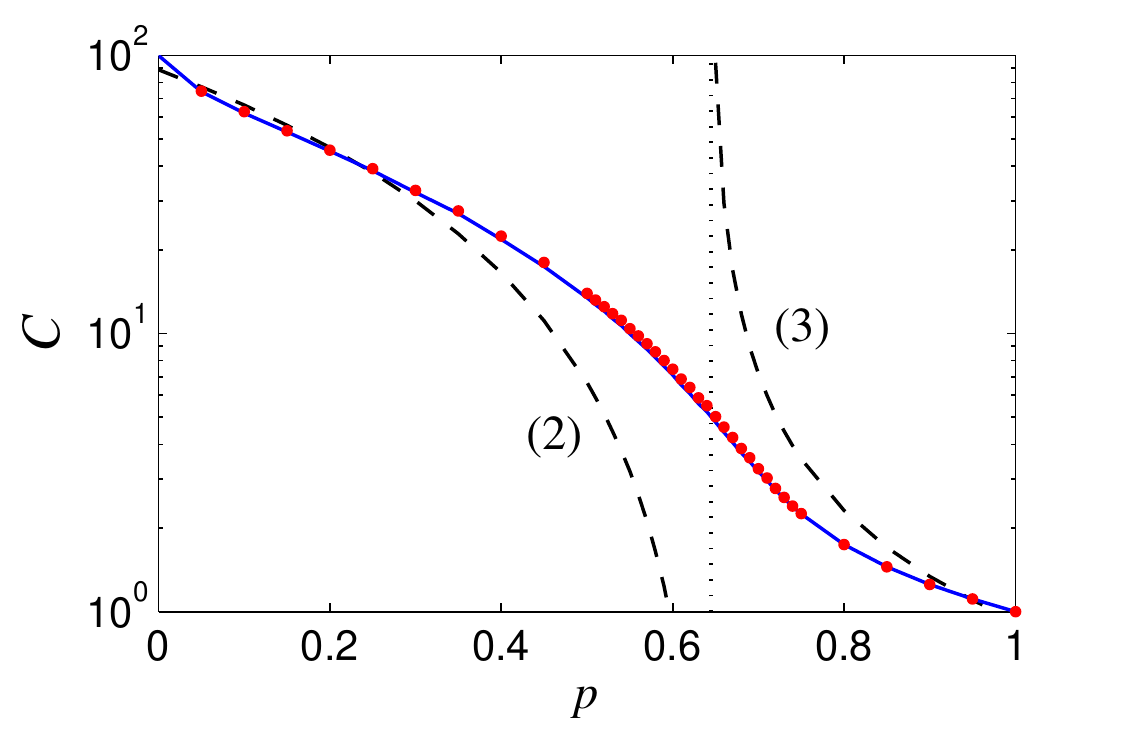}
\caption{(Color online) Dependence of the commute time on the fraction $p$ of congestible roads, plotted for $L = 50$.  The solid (blue) line shows $\copt$, and the (red) dots are $\ceq$.  The dashed lines show the analytical results of Eqs.\ (\ref{eq:Cpsmall}) and (\ref{eq:Cplarge}), respectively, for $L \rightarrow \infty$.  The dotted vertical line indicates $p = p_c$.}
\label{fig:scaling_with_p}
\end{figure}

On the other hand, at $p > p_c$, there are many parallel pathways for traversing the lattice that use only congestible roads, as depicted in Fig.\ \ref{fig:clusters}(b).  Each such pathway takes only a small fraction $x' \sim \xipd/L$ of the total current, while paths that traverse constant roads are completely abandoned.  Since the total commute time along any given percolating pathway is proportional to the commute time $c(x') = x'$ on a given congestible road multiplied by the total path length $\sim L$, one can say that the typical average time $C \sim x' L$.  In other words, 
\be 
C \sim \xipd \sim |p - p_c|^{-\nupd}
\label{eq:Cplarge}
\ee 
at $p > p_c$.  Equations (\ref{eq:Cpsmall}) and (\ref{eq:Cplarge}) are shown together with numerical results for $C$ in Fig.\ \ref{fig:scaling_with_p}.

\begin{figure}[htb]
\centering
\includegraphics[width=.9\columnwidth]{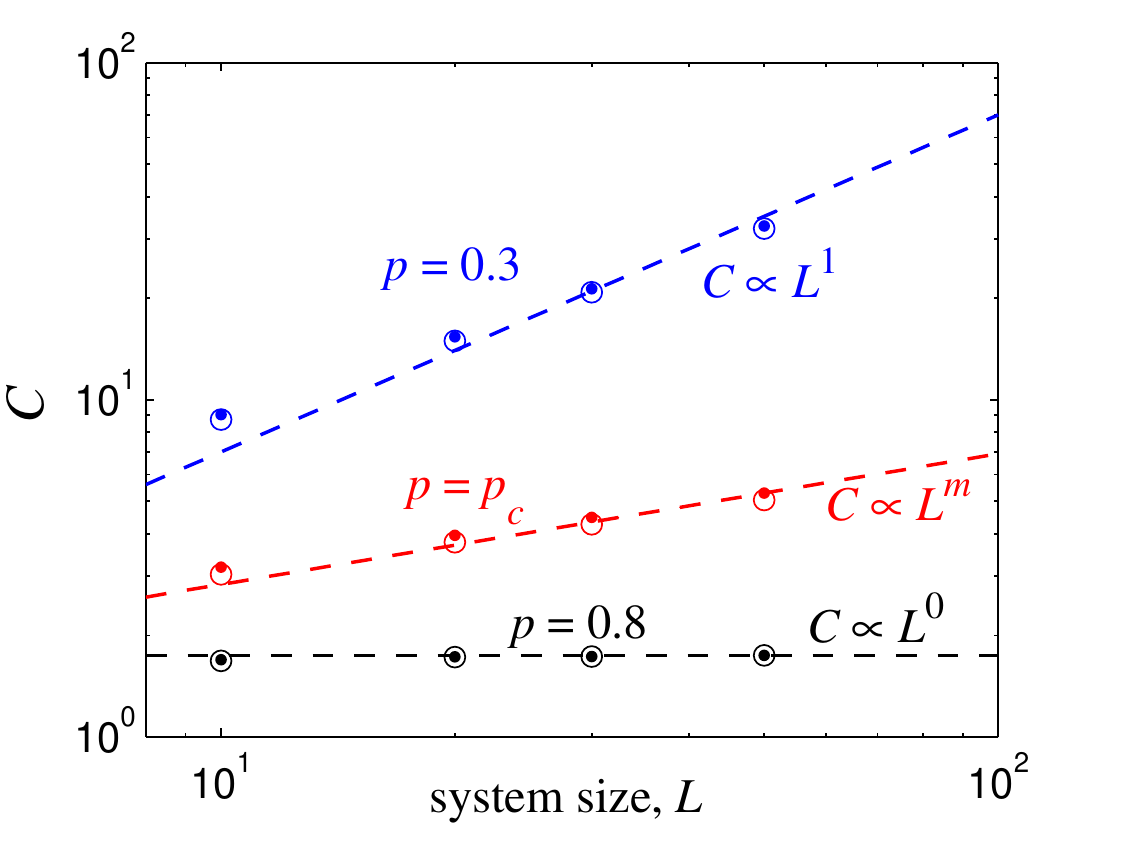}
\caption{(Color online) Scaling of the average commute time $C$ with the system size $L$.  Different curves are labelled by their corresponding value of $p$.  At $p < p_c$, the commute time scales as $C \propto L^1$, and at $p > p_c$ the commute time scales as $C \propto L^{0}$.  Precisely at $p = p_c$, the commute time follows $C \propto L^{m}$, with $m \approx 0.388$.  Filled dots denote $\ceq$, and open circles are $\copt$.}
\label{fig:scaling_with_L}
\end{figure}

One can notice that Eqs.\ (\ref{eq:Cpsmall}) and (\ref{eq:Cplarge}) represent very different behavior: at $p < p_c$ the commute time scales extensively with the system size, $C \propto L^{1}$, while at $p > p_c$ the commute time becomes independent of system size, $C \propto L^0$.  Exactly at the threshold, $p = p_c$, one can expect the commute time to scale as some nontrivial power of the system size: $C \propto L^m$.  An estimate for this exponent $m$ can be obtained by equating Eqs.\ (\ref{eq:Cpsmall}) and (\ref{eq:Cplarge}) and solving for the corresponding value of $|p-p_c|$ at which the two relations cross over to each other.  This procedure gives
\be 
C(p = p_c) \sim L^{\nupd/(\nupl + \nupd)},
\ee 
or $m = \nupd/(\nupl + \nupd) \approx 0.388$.  This analytical estimate for $m$ is consistent with numerical results, as shown in Fig.\ \ref{fig:scaling_with_L}.  An independent fitting of the data in Fig.\ \ref{fig:scaling_with_L} gives $m \approx 0.31$.  This same scaling relation $C \propto L^{m}$ describes both the equilibrium and optimum. (Indeed, general theorems have shown that $\ceq$ and $\copt$ cannot differ by more than a constant numeric factor \cite{roughgarden_how_2002}.  For networks with linear cost functions, $\ceq/copt \leq 4/3$.)

\begin{figure}[htb]
\centering
\includegraphics[width=\columnwidth]{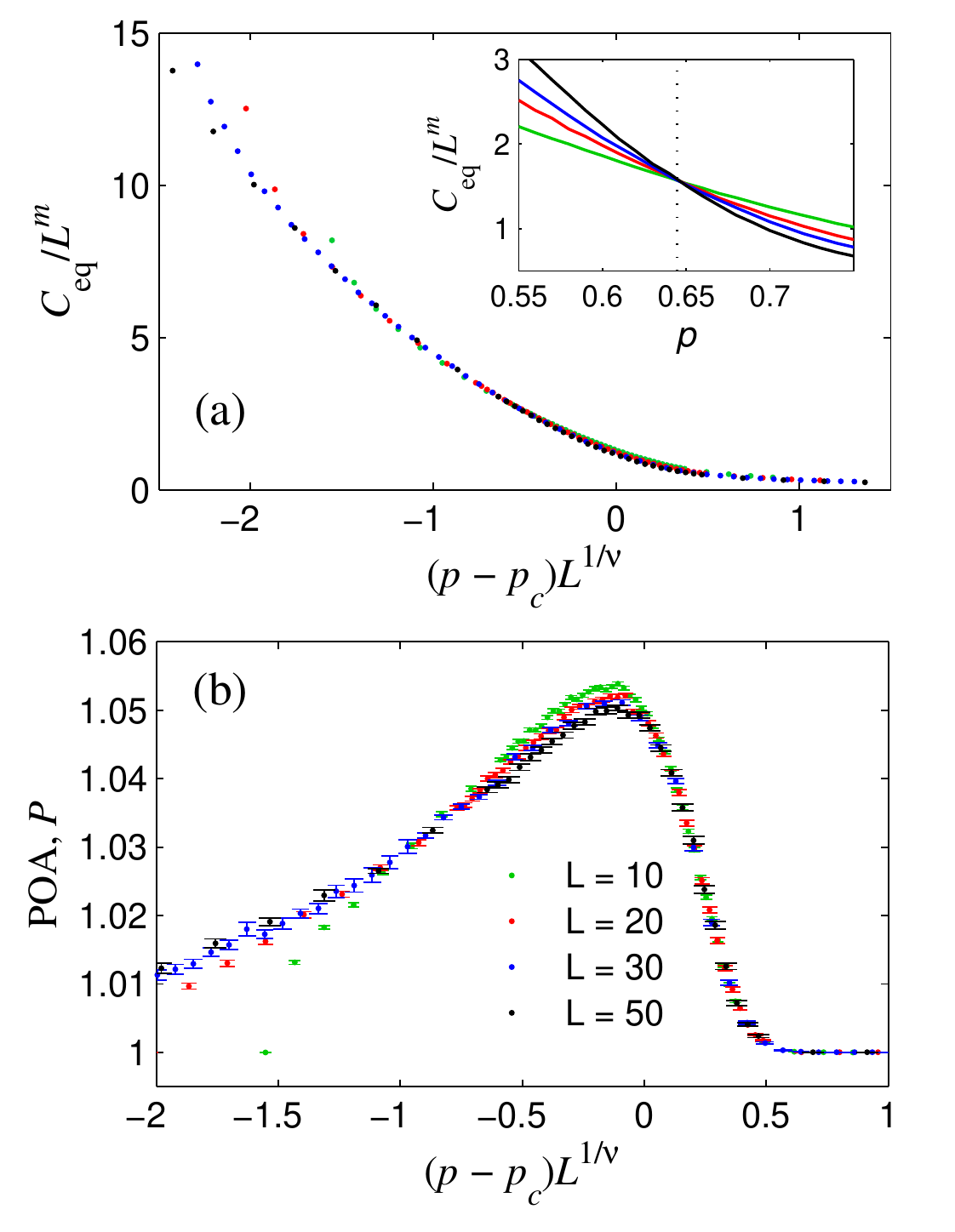}
\caption{(Color online) Critical scaling.  (a) Inset: The scaled equilibrium commute time $\ceq/L^m$ near the percolation threshold (vertical dotted line).  Main figure: $\ceq/L^{m}$ can be parameterized for all system sizes by $(p-p_c) L^{1/\nu}$, with $\nu \approx 2.6$.  $\copt$ (not shown) can be similarly scaled. (b) Near the percolation threshold, different curves for the POA also collapse when plotted as a function of $(p-p_c) L^{1/\nu}$.}
\label{fig:scaling}
\end{figure}

For finite values of the system size, and for $p$ close to the percolation threshold, it should be possible to write the scaled commute time $C/L^m$ using a critical scaling form.  Such scaling usually takes the form $F = f(L/\xi)$, where $F$ is some system property and $\xi \propto |p - p_c|^{-\nu}$ is the correlation length, so that $F$ is a function only of $L/\xi$, or equilvalently of $(L/\xi)^{1/\nu} = (p-p_c)L^{1/\nu}$.  Such scaling is indeed possible here, as demonstrated explicitly in Fig.\ \ref{fig:scaling}(a).  In this figure the scaled cost $C/L^m$ is shown to be a function only of the combination $(p - p_c)L^{1/\nu}$, with $\nu \approx 2.62 \pm 0.2$ determined by best fit of the data collapse\cite{bhattacharjee_measure_2001}.  Fig.\ \ref{fig:scaling}(b) demonstrates that this same value of $\nu$ produces good gollapse of the curves from Fig.\ \ref{fig:POA} for the POA.  As mentioned above, such scaling suggests that for $L \rightarrow \infty$ the POA peak converges precisely to $p = p_c$.

Of course, the problem being considered here has two separate correlation lengths, so that the fitted exponent $\nu$ should in fact be a linear combination of the two critical exponents $\nupl$ and $\nupd$.  One reasonable expectation is that the relevant scaling form is $C/L^{m} = f(L/L_c)$, where $L_c \sim \xipd \xipl$ is the system size for which Eqs.\ (\ref{eq:Cpsmall}) and (\ref{eq:Cplarge}) become equal.  In other words, $L_c$ represents the system size below which the system cannot be unambiguously described as percolating or non-percolating, in terms of its average commute time.  This hypothesis leads to the conclusion that $\nu = \nupl + \nupd \approx 2.83$, which is consistent with the result presented above.  Future works can validate this hypothesis explicitly by studying systems with varying aspect ratio.  

\section{Conclusion}
\label{sec:conclusion}

In summary, this paper has introduced a simple model of random congestible networks and demonstrated a clear and previously unnoticed connection between percolation and self-interest-driven inefficiency.  A number of generalizations and extensions of the model deserve further exploration, including an extension to other network topologies \cite{valiant_braesss_2010, hata_advection_2014}, and to the case of nonlinear cost functions.  For highly nonlinear cost functions one can, in general, expect a significantly larger value of the POA at the percolation threshold\footnote{Generally, a network whose cost functions are monotonically increasing and involve a power law of degree no larger than $\alpha$ have $P \leq (\alpha+1)^{1+1/\alpha}/[(\alpha+1)^{1+1/\alpha} - \alpha]$ \cite{roughgarden_price_2003}}.  Cost functions that include a critical ``jamming" density, as are realistic for actual highways \cite{may_traffic_1990}, may give rise to a diverging POA.

More generally, this work hints at the possibility of a deeper connection between POA and percolation that may go well beyond simple Pigou-type models.  Such a connection seems to provide a novel and interesting playground for statistical mechanics, and has the potential to provide invaluable understanding for efforts to mitigate network congestion effects across a wide set of disciplines.

\acknowledgments
I am grateful to A.\ Nahum, V.\ Sokolov, K.\ A.\ Matveev, T.\ Roughgarden, A.\ Glatz, and R.\ Brierley for helpful discussions, and to W.\ DeGottardi, G.\ Guzman-Verri, and A.\ Lopez-Bezanilla for their comments on the manuscript.
Work at Argonne National Laboratory was supported by the U.S. Department of Energy, Office of Science under contract no. DE-AC02-06CH11357.

\appendix

\pagebreak 

\begin{widetext}

\section{POA in a different lattice type}
\label{app:3D}

As mentioned in the main text, one natural test of the generality of the claim that the POA is maximized at $p = p_c$ is to analyze flow through a different type of lattice.  Here I show results for the three-dimensional (3D) body-centered cubic (BCC) lattice, which can be thought of as the 3D generalization of the lattice in Fig.\ \ref{fig:schematics}(b).  The BCC lattice has a significantly smaller percolation threshold $p_c \approx 0.2873$ due to its twice-larger coordination number.  The POA as a function of $p$ for this lattice is plotted in Fig.\ \ref{fig:3D}.  As one can see, the behavior of $P(p)$ is qualitatively similar to that of Fig.\ \ref{fig:POA}, but with the maximum shifted to the new percolation threshold.

\begin{figure}[htb]
\centering
\includegraphics[width=0.45 \textwidth]{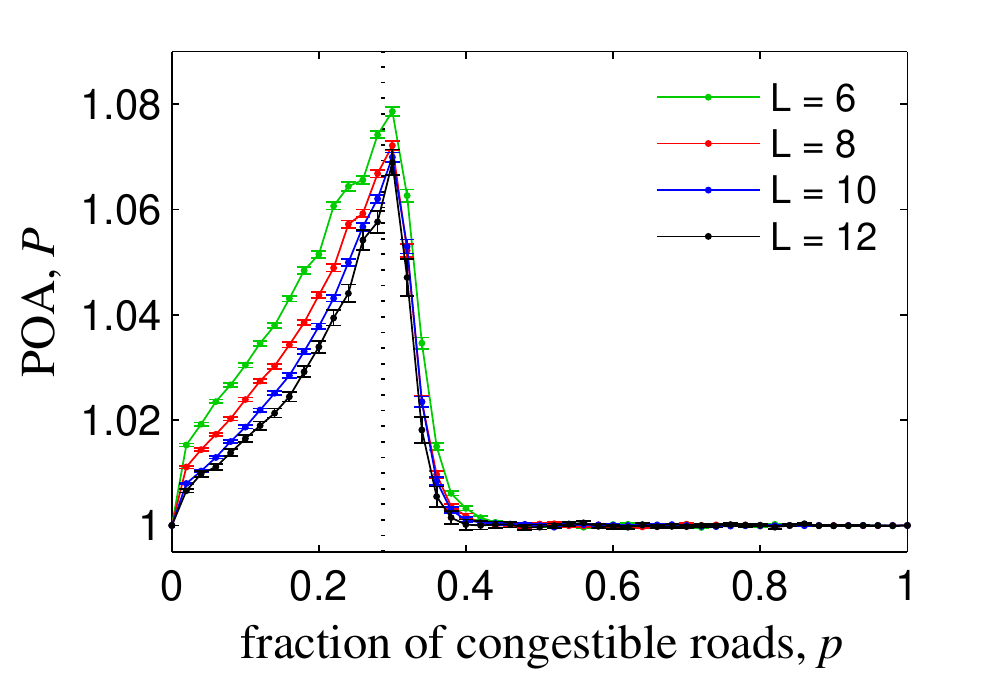}
\caption{(Color online) POA for flow through a 3D BCC lattice with randomly-placed congestible and constant roads, plotted for different system sizes.  The vertical dotted line shows the lattice percolation threshold, $p_c \approx 0.2873$.}
\label{fig:3D}
\end{figure}

\section{Traffic density maps}
\label{app:maps}

Figure \ref{fig:currents} of the main text provides a visualization of the traffic through a particular realization of the network by showing which roads are used in the equilibrium and optimum situations.  Here I provide an additional visualization of the traffic density.  

In Fig.\ \ref{fig:maps} are shown the equilibrium and optimum traffic densities $x$ for each road of the same random network depicted in Fig.\ \ref{fig:currents}.  In addition to having smaller ``holes" in the traffic pattern (as illustrated in Fig.\ \ref{fig:currents}), the optimum also has a more even distribution of the traffic flow as compared to the equilibrium.

\begin{figure}[htb]
\centering
\includegraphics[width=0.7 \textwidth]{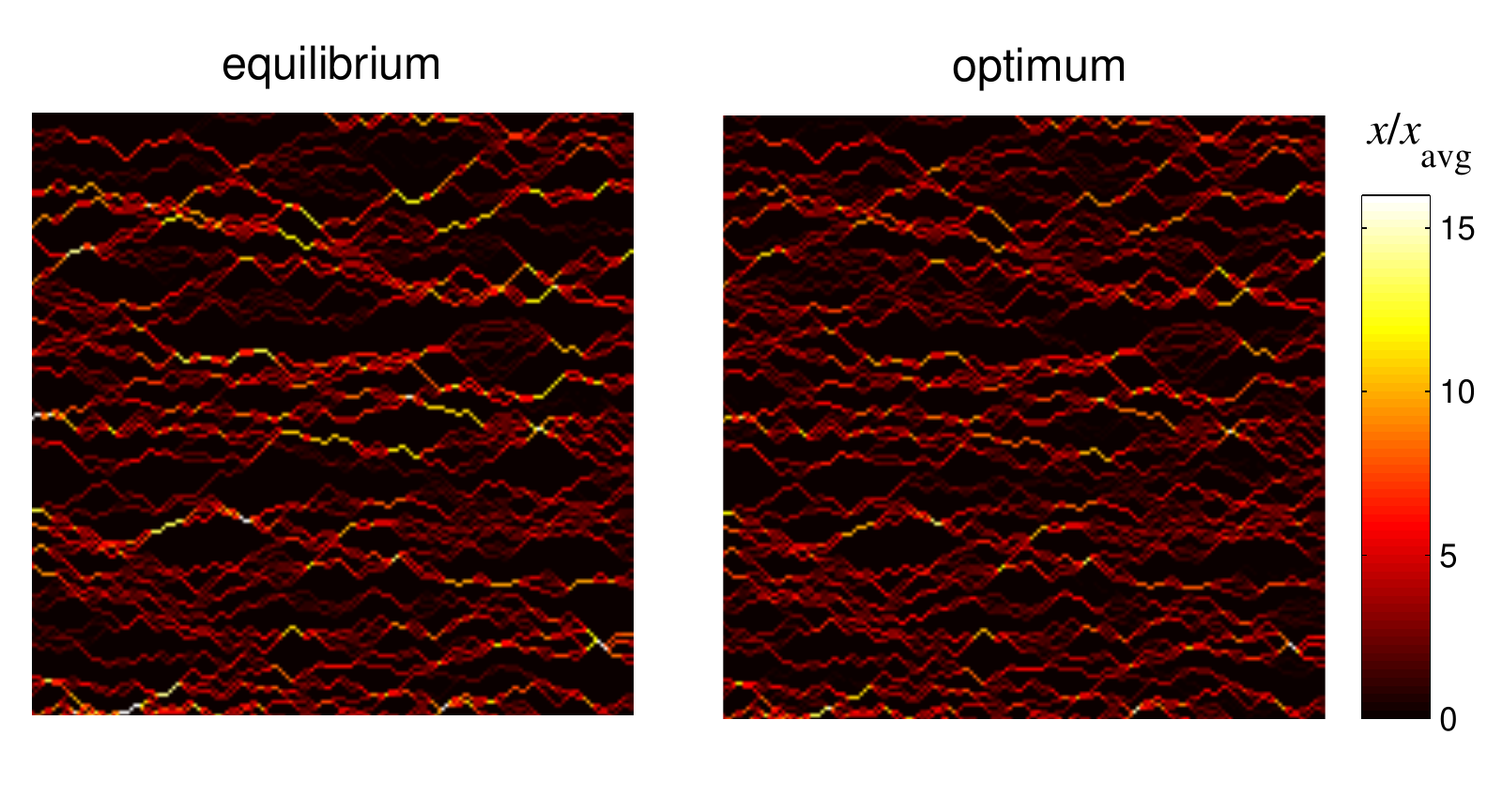}
\caption{(Color online) Traffic density in a network close to the percolation threshold, $p \approx p_c$.  The color indicates the traffic on a particular road, normalized to the system-averaged traffic per road, $x_\text{avg} = 1/(2L)$.  The equilibrium configuration (left) generally has its traffic concentrated on fewer roads, while the optimum (right) has a more even distribution.}
\label{fig:maps}
\end{figure}

\end{widetext}

\bibliography{pigou}

\begin{thebibliography}{33}%
\makeatletter
\providecommand \@ifxundefined [1]{%
 \@ifx{#1\undefined}
}%
\providecommand \@ifnum [1]{%
 \ifnum #1\expandafter \@firstoftwo
 \else \expandafter \@secondoftwo
 \fi
}%
\providecommand \@ifx [1]{%
 \ifx #1\expandafter \@firstoftwo
 \else \expandafter \@secondoftwo
 \fi
}%
\providecommand \natexlab [1]{#1}%
\providecommand \enquote  [1]{``#1''}%
\providecommand \bibnamefont  [1]{#1}%
\providecommand \bibfnamefont [1]{#1}%
\providecommand \citenamefont [1]{#1}%
\providecommand \href@noop [0]{\@secondoftwo}%
\providecommand \href [0]{\begingroup \@sanitize@url \@href}%
\providecommand \@href[1]{\@@startlink{#1}\@@href}%
\providecommand \@@href[1]{\endgroup#1\@@endlink}%
\providecommand \@sanitize@url [0]{\catcode `\\12\catcode `\$12\catcode
  `\&12\catcode `\#12\catcode `\^12\catcode `\_12\catcode `\%12\relax}%
\providecommand \@@startlink[1]{}%
\providecommand \@@endlink[0]{}%
\providecommand \url  [0]{\begingroup\@sanitize@url \@url }%
\providecommand \@url [1]{\endgroup\@href {#1}{\urlprefix }}%
\providecommand \urlprefix  [0]{URL }%
\providecommand \Eprint [0]{\href }%
\providecommand \doibase [0]{http://dx.doi.org/}%
\providecommand \selectlanguage [0]{\@gobble}%
\providecommand \bibinfo  [0]{\@secondoftwo}%
\providecommand \bibfield  [0]{\@secondoftwo}%
\providecommand \translation [1]{[#1]}%
\providecommand \BibitemOpen [0]{}%
\providecommand \bibitemStop [0]{}%
\providecommand \bibitemNoStop [0]{.\EOS\space}%
\providecommand \EOS [0]{\spacefactor3000\relax}%
\providecommand \BibitemShut  [1]{\csname bibitem#1\endcsname}%
\let\auto@bib@innerbib\@empty
\bibitem [{\citenamefont {Roughgarden}(2005)}]{roughgarden_selfish_2005}%
  \BibitemOpen
  \bibfield  {author} {\bibinfo {author} {\bibfnamefont {T.}~\bibnamefont
  {Roughgarden}},\ }\href@noop {} {\emph {\bibinfo {title} {Selfish routing and
  the price of anarchy}}}\ (\bibinfo  {publisher} {{MIT} press},\ \bibinfo
  {year} {2005})\BibitemShut {NoStop}%
\bibitem [{\citenamefont {Braess}(1968)}]{braess_uber_1968}%
  \BibitemOpen
  \bibfield  {author} {\bibinfo {author} {\bibfnamefont {D.}~\bibnamefont
  {Braess}},\ }\href {\doibase 10.1007/BF01918335} {\bibfield  {journal}
  {\bibinfo  {journal} {Zeitschrift für Operations-Research}\ }\textbf
  {\bibinfo {volume} {12}},\ \bibinfo {pages} {258} (\bibinfo {year}
  {1968})}\BibitemShut {NoStop}%
\bibitem [{\citenamefont {Wardrop}(1952)}]{wardrop_road_1952}%
  \BibitemOpen
  \bibfield  {author} {\bibinfo {author} {\bibfnamefont {J.~G.}\ \bibnamefont
  {Wardrop}},\ }in\ \href@noop {} {\emph {\bibinfo {booktitle} {{ICE}
  Proceedings: Engineering Divisions}}},\ Vol.~\bibinfo {volume} {1}\ (\bibinfo
   {publisher} {Thomas Telford},\ \bibinfo {year} {1952})\ pp.\ \bibinfo
  {pages} {325--362}\BibitemShut {NoStop}%
\bibitem [{\citenamefont {Youn}\ \emph {et~al.}(2008)\citenamefont {Youn},
  \citenamefont {Gastner},\ and\ \citenamefont {Jeong}}]{youn_price_2008}%
  \BibitemOpen
  \bibfield  {author} {\bibinfo {author} {\bibfnamefont {H.}~\bibnamefont
  {Youn}}, \bibinfo {author} {\bibfnamefont {M.~T.}\ \bibnamefont {Gastner}}, \
  and\ \bibinfo {author} {\bibfnamefont {H.}~\bibnamefont {Jeong}},\ }\href
  {\doibase 10.1103/PhysRevLett.101.128701} {\bibfield  {journal} {\bibinfo
  {journal} {Physical Review Letters}\ }\textbf {\bibinfo {volume} {101}},\
  \bibinfo {pages} {128701} (\bibinfo {year} {2008})}\BibitemShut {NoStop}%
\bibitem [{\citenamefont {Zhu}\ \emph {et~al.}(2010)\citenamefont {Zhu},
  \citenamefont {Levinson},\ and\ \citenamefont {Liu}}]{zhu_measuring_2010}%
  \BibitemOpen
  \bibfield  {author} {\bibinfo {author} {\bibfnamefont {S.}~\bibnamefont
  {Zhu}}, \bibinfo {author} {\bibfnamefont {D.}~\bibnamefont {Levinson}}, \
  and\ \bibinfo {author} {\bibfnamefont {H.}~\bibnamefont {Liu}},\ }in\
  \href@noop {} {\emph {\bibinfo {booktitle} {Transportation Research Board
  89th Annual Meeting Compendium of Papers}}},\ Vol.\ \bibinfo {volume} {1005}\
  (\bibinfo  {publisher} {Citeseer},\ \bibinfo {year} {2010})\ pp.\ \bibinfo
  {pages} {10--2298}\BibitemShut {NoStop}%
\bibitem [{\citenamefont {Witthaut}\ and\ \citenamefont
  {Timme}(2012)}]{witthaut_braesss_2012}%
  \BibitemOpen
  \bibfield  {author} {\bibinfo {author} {\bibfnamefont {D.}~\bibnamefont
  {Witthaut}}\ and\ \bibinfo {author} {\bibfnamefont {M.}~\bibnamefont
  {Timme}},\ }\href {\doibase 10.1088/1367-2630/14/8/083036} {\bibfield
  {journal} {\bibinfo  {journal} {New Journal of Physics}\ }\textbf {\bibinfo
  {volume} {14}},\ \bibinfo {pages} {083036} (\bibinfo {year}
  {2012})}\BibitemShut {NoStop}%
\bibitem [{\citenamefont {Knight}\ and\ \citenamefont
  {Harper}(2013)}]{knight_selfish_2013}%
  \BibitemOpen
  \bibfield  {author} {\bibinfo {author} {\bibfnamefont {V.~A.}\ \bibnamefont
  {Knight}}\ and\ \bibinfo {author} {\bibfnamefont {P.~R.}\ \bibnamefont
  {Harper}},\ }\href {\doibase 10.1016/j.ejor.2013.04.003} {\bibfield
  {journal} {\bibinfo  {journal} {European Journal of Operational Research}\
  }\textbf {\bibinfo {volume} {230}},\ \bibinfo {pages} {122} (\bibinfo {year}
  {2013})}\BibitemShut {NoStop}%
\bibitem [{\citenamefont {Skinner}(2010)}]{skinner_price_2010}%
  \BibitemOpen
  \bibfield  {author} {\bibinfo {author} {\bibfnamefont {B.}~\bibnamefont
  {Skinner}},\ }\href {http://dx.doi.org/10.2202/1559-0410.1217} {\bibfield
  {journal} {\bibinfo  {journal} {Journal of Quantitative Analysis in Sports}\
  }\textbf {\bibinfo {volume} {6}},\ \bibinfo {pages} {Article 3} (\bibinfo
  {year} {2010})}\BibitemShut {NoStop}%
\bibitem [{\citenamefont {Skinner}\ and\ \citenamefont
  {Carlin}(2013)}]{skinner_price_2013}%
  \BibitemOpen
  \bibfield  {author} {\bibinfo {author} {\bibfnamefont {B.}~\bibnamefont
  {Skinner}}\ and\ \bibinfo {author} {\bibfnamefont {B.}~\bibnamefont
  {Carlin}},\ }\href {\doibase 10.1111/j.1740-9713.2013.00662.x} {\bibfield
  {journal} {\bibinfo  {journal} {Significance}\ }\textbf {\bibinfo {volume}
  {10}},\ \bibinfo {pages} {25} (\bibinfo {year} {2013})}\BibitemShut {NoStop}%
\bibitem [{\citenamefont {Valiant}\ and\ \citenamefont
  {Roughgarden}(2010)}]{valiant_braesss_2010}%
  \BibitemOpen
  \bibfield  {author} {\bibinfo {author} {\bibfnamefont {G.}~\bibnamefont
  {Valiant}}\ and\ \bibinfo {author} {\bibfnamefont {T.}~\bibnamefont
  {Roughgarden}},\ }\href {\doibase 10.1002/rsa.20325} {\bibfield  {journal}
  {\bibinfo  {journal} {Random Structures \& Algorithms}\ }\textbf {\bibinfo
  {volume} {37}},\ \bibinfo {pages} {495} (\bibinfo {year} {2010})}\BibitemShut
  {NoStop}%
\bibitem [{\citenamefont {Roughgarden}\ and\ \citenamefont
  {Tardos}(2002)}]{roughgarden_how_2002}%
  \BibitemOpen
  \bibfield  {author} {\bibinfo {author} {\bibfnamefont {T.}~\bibnamefont
  {Roughgarden}}\ and\ \bibinfo {author} {\bibfnamefont {v.}~\bibnamefont
  {Tardos}},\ }\href {\doibase 10.1145/506147.506153} {\bibfield  {journal}
  {\bibinfo  {journal} {J. {ACM}}\ }\textbf {\bibinfo {volume} {49}},\ \bibinfo
  {pages} {236} (\bibinfo {year} {2002})}\BibitemShut {NoStop}%
\bibitem [{\citenamefont {Roughgarden}(2003)}]{roughgarden_price_2003}%
  \BibitemOpen
  \bibfield  {author} {\bibinfo {author} {\bibfnamefont {T.}~\bibnamefont
  {Roughgarden}},\ }\href {\doibase 10.1016/S0022-0000(03)00044-8} {\bibfield
  {journal} {\bibinfo  {journal} {Journal of Computer and System Sciences}\
  }\bibinfo {series} {Special Issue on {STOC} 2002},\ \textbf {\bibinfo
  {volume} {67}},\ \bibinfo {pages} {341} (\bibinfo {year} {2003})}\BibitemShut
  {NoStop}%
\bibitem [{\citenamefont {Tsekeris}\ and\ \citenamefont
  {Voß}(2009)}]{tsekeris_design_2009}%
  \BibitemOpen
  \bibfield  {author} {\bibinfo {author} {\bibfnamefont {T.}~\bibnamefont
  {Tsekeris}}\ and\ \bibinfo {author} {\bibfnamefont {S.}~\bibnamefont
  {Voß}},\ }\href {\doibase 10.1007/s11066-008-9024-z} {\bibfield  {journal}
  {\bibinfo  {journal} {{NETNOMICS}: Economic Research and Electronic
  Networking}\ }\textbf {\bibinfo {volume} {10}},\ \bibinfo {pages} {5}
  (\bibinfo {year} {2009})}\BibitemShut {NoStop}%
\bibitem [{\citenamefont {Fisch}\ and\ \citenamefont
  {Harris}(1978)}]{fisch_critical_1978}%
  \BibitemOpen
  \bibfield  {author} {\bibinfo {author} {\bibfnamefont {R.}~\bibnamefont
  {Fisch}}\ and\ \bibinfo {author} {\bibfnamefont {A.~B.}\ \bibnamefont
  {Harris}},\ }\href {\doibase 10.1103/PhysRevB.18.416} {\bibfield  {journal}
  {\bibinfo  {journal} {Physical Review B}\ }\textbf {\bibinfo {volume} {18}},\
  \bibinfo {pages} {416} (\bibinfo {year} {1978})}\BibitemShut {NoStop}%
\bibitem [{\citenamefont {Skal}\ and\ \citenamefont
  {Shklovskii}(1975)}]{skal_topology_1975}%
  \BibitemOpen
  \bibfield  {author} {\bibinfo {author} {\bibfnamefont {A.}~\bibnamefont
  {Skal}}\ and\ \bibinfo {author} {\bibfnamefont {B.}~\bibnamefont
  {Shklovskii}},\ }\href@noop {} {\bibfield  {journal} {\bibinfo  {journal}
  {Sov. Phys. Semicond.-{USSR}}\ }\textbf {\bibinfo {volume} {8}},\ \bibinfo
  {pages} {1029} (\bibinfo {year} {1975})}\BibitemShut {NoStop}%
\bibitem [{\citenamefont {Nguen}\ \emph {et~al.}(1985)\citenamefont {Nguen},
  \citenamefont {Spivak},\ and\ \citenamefont
  {Shklovksii}}]{nguen_tunnel_1985}%
  \BibitemOpen
  \bibfield  {author} {\bibinfo {author} {\bibfnamefont {V.~L.}\ \bibnamefont
  {Nguen}}, \bibinfo {author} {\bibfnamefont {B.~Z.}\ \bibnamefont {Spivak}}, \
  and\ \bibinfo {author} {\bibfnamefont {B.~I.}\ \bibnamefont {Shklovksii}},\
  }\href@noop {} {\bibfield  {journal} {\bibinfo  {journal} {Zh. Eksp. Teor.
  Fiz}\ }\textbf {\bibinfo {volume} {89}},\ \bibinfo {pages} {1770} (\bibinfo
  {year} {1985})},\ \bibinfo {note} {[{JETP} 62, 1021, (1985)]}\BibitemShut
  {NoStop}%
\bibitem [{\citenamefont {Moore}\ and\ \citenamefont
  {Newman}(2000)}]{moore_epidemics_2000}%
  \BibitemOpen
  \bibfield  {author} {\bibinfo {author} {\bibfnamefont {C.}~\bibnamefont
  {Moore}}\ and\ \bibinfo {author} {\bibfnamefont {M.~E.~J.}\ \bibnamefont
  {Newman}},\ }\href {\doibase 10.1103/PhysRevE.61.5678} {\bibfield  {journal}
  {\bibinfo  {journal} {Physical Review E}\ }\textbf {\bibinfo {volume} {61}},\
  \bibinfo {pages} {5678} (\bibinfo {year} {2000})}\BibitemShut {NoStop}%
\bibitem [{\citenamefont {Sander}\ \emph {et~al.}(2002)\citenamefont {Sander},
  \citenamefont {Warren}, \citenamefont {Sokolov}, \citenamefont {Simon},\ and\
  \citenamefont {Koopman}}]{sander_percolation_2002}%
  \BibitemOpen
  \bibfield  {author} {\bibinfo {author} {\bibfnamefont {L.~M.}\ \bibnamefont
  {Sander}}, \bibinfo {author} {\bibfnamefont {C.~P.}\ \bibnamefont {Warren}},
  \bibinfo {author} {\bibfnamefont {I.~M.}\ \bibnamefont {Sokolov}}, \bibinfo
  {author} {\bibfnamefont {C.}~\bibnamefont {Simon}}, \ and\ \bibinfo {author}
  {\bibfnamefont {J.}~\bibnamefont {Koopman}},\ }\href {\doibase
  10.1016/S0025-5564(02)00117-7} {\bibfield  {journal} {\bibinfo  {journal}
  {Mathematical Biosciences}\ }\textbf {\bibinfo {volume} {180}},\ \bibinfo
  {pages} {293} (\bibinfo {year} {2002})}\BibitemShut {NoStop}%
\bibitem [{\citenamefont {Broedersz}\ \emph {et~al.}(2011)\citenamefont
  {Broedersz}, \citenamefont {Mao}, \citenamefont {Lubensky},\ and\
  \citenamefont {MacKintosh}}]{broedersz_criticality_2011}%
  \BibitemOpen
  \bibfield  {author} {\bibinfo {author} {\bibfnamefont {C.~P.}\ \bibnamefont
  {Broedersz}}, \bibinfo {author} {\bibfnamefont {X.}~\bibnamefont {Mao}},
  \bibinfo {author} {\bibfnamefont {T.~C.}\ \bibnamefont {Lubensky}}, \ and\
  \bibinfo {author} {\bibfnamefont {F.~C.}\ \bibnamefont {MacKintosh}},\ }\href
  {\doibase 10.1038/nphys2127} {\bibfield  {journal} {\bibinfo  {journal}
  {Nature Physics}\ }\textbf {\bibinfo {volume} {7}},\ \bibinfo {pages} {983}
  (\bibinfo {year} {2011})}\BibitemShut {NoStop}%
\bibitem [{\citenamefont {Chalker}\ and\ \citenamefont
  {Coddington}(1988)}]{chalker_percolation_1988}%
  \BibitemOpen
  \bibfield  {author} {\bibinfo {author} {\bibfnamefont {J.~T.}\ \bibnamefont
  {Chalker}}\ and\ \bibinfo {author} {\bibfnamefont {P.~D.}\ \bibnamefont
  {Coddington}},\ }\href {\doibase 10.1088/0022-3719/21/14/008} {\bibfield
  {journal} {\bibinfo  {journal} {Journal of Physics C: Solid State Physics}\
  }\textbf {\bibinfo {volume} {21}},\ \bibinfo {pages} {2665} (\bibinfo {year}
  {1988})}\BibitemShut {NoStop}%
\bibitem [{Note1()}]{Note1}%
  \BibitemOpen
  \bibinfo {note} {Such language may become particularly relevant with the
  further development of self-driving cars\cite
  {markoff_google_2010}.}\BibitemShut {Stop}%
\bibitem [{\citenamefont {Pigou}(1924)}]{pigou_economics_1924}%
  \BibitemOpen
  \bibfield  {author} {\bibinfo {author} {\bibfnamefont {A.~C.}\ \bibnamefont
  {Pigou}},\ }\href@noop {} {\emph {\bibinfo {title} {The economics of
  welfare}}}\ (\bibinfo  {publisher} {Transaction Publishers},\ \bibinfo {year}
  {1924})\BibitemShut {NoStop}%
\bibitem [{\citenamefont {Dasgupta}\ \emph {et~al.}(2008)\citenamefont
  {Dasgupta}, \citenamefont {Papadimitriou},\ and\ \citenamefont
  {Vazirani}}]{dasgupta_algorithms_2008}%
  \BibitemOpen
  \bibfield  {author} {\bibinfo {author} {\bibfnamefont {S.}~\bibnamefont
  {Dasgupta}}, \bibinfo {author} {\bibfnamefont {C.~H.}\ \bibnamefont
  {Papadimitriou}}, \ and\ \bibinfo {author} {\bibfnamefont {U.}~\bibnamefont
  {Vazirani}},\ }\href@noop {} {\emph {\bibinfo {title} {Algorithms}}},\
  \bibinfo {edition} {1st}\ ed.\ (\bibinfo  {publisher} {{McGraw}-Hill, Inc.},\
  \bibinfo {address} {New York, {NY}, {USA}},\ \bibinfo {year}
  {2008})\BibitemShut {NoStop}%
\bibitem [{\citenamefont {Shklovskii}\ and\ \citenamefont
  {Efros}(1984)}]{shklovskii_electronic_1984}%
  \BibitemOpen
  \bibfield  {author} {\bibinfo {author} {\bibfnamefont {B.~I.}\ \bibnamefont
  {Shklovskii}}\ and\ \bibinfo {author} {\bibfnamefont {A.~L.}\ \bibnamefont
  {Efros}},\ }\href@noop {} {\emph {\bibinfo {title} {Electronic Properties of
  Doped Semiconductors}}}\ (\bibinfo  {publisher} {Springer-Verlag},\ \bibinfo
  {address} {New York},\ \bibinfo {year} {1984})\BibitemShut {NoStop}%
\bibitem [{\citenamefont {Black}\ and\ \citenamefont
  {Saltzer}(1960)}]{black_resistive_1960}%
  \BibitemOpen
  \bibfield  {author} {\bibinfo {author} {\bibfnamefont {W.~L.}\ \bibnamefont
  {Black}}\ and\ \bibinfo {author} {\bibfnamefont {J.~H.}\ \bibnamefont
  {Saltzer}},\ }\href
  {http://web.mit.edu/saltzer/www/publications/RDNT.ocr.pdf} {\bibfield
  {journal} {\bibinfo  {journal} {Tech Engineering News}\ }\textbf {\bibinfo
  {volume} {{XLI}}},\ \bibinfo {pages} {18} (\bibinfo {year}
  {1960})}\BibitemShut {NoStop}%
\bibitem [{\citenamefont {Jensen}(1999)}]{jensen_low-density_1999}%
  \BibitemOpen
  \bibfield  {author} {\bibinfo {author} {\bibfnamefont {I.}~\bibnamefont
  {Jensen}},\ }\href {\doibase 10.1088/0305-4470/32/28/304} {\bibfield
  {journal} {\bibinfo  {journal} {Journal of Physics A: Mathematical and
  General}\ }\textbf {\bibinfo {volume} {32}},\ \bibinfo {pages} {5233}
  (\bibinfo {year} {1999})}\BibitemShut {NoStop}%
\bibitem [{\citenamefont {Family}(1982)}]{family_relation_1982}%
  \BibitemOpen
  \bibfield  {author} {\bibinfo {author} {\bibfnamefont {F.}~\bibnamefont
  {Family}},\ }\href {\doibase 10.1088/0305-4470/15/11/003} {\bibfield
  {journal} {\bibinfo  {journal} {Journal of Physics A: Mathematical and
  General}\ }\textbf {\bibinfo {volume} {15}},\ \bibinfo {pages} {L583}
  (\bibinfo {year} {1982})}\BibitemShut {NoStop}%
\bibitem [{\citenamefont {Redner}(1982)}]{redner_directed_1982}%
  \BibitemOpen
  \bibfield  {author} {\bibinfo {author} {\bibfnamefont {S.}~\bibnamefont
  {Redner}},\ }\href {\doibase 10.1103/PhysRevB.25.3242} {\bibfield  {journal}
  {\bibinfo  {journal} {Physical Review B}\ }\textbf {\bibinfo {volume} {25}},\
  \bibinfo {pages} {3242} (\bibinfo {year} {1982})}\BibitemShut {NoStop}%
\bibitem [{\citenamefont {Bhattacharjee}\ and\ \citenamefont
  {Seno}(2001)}]{bhattacharjee_measure_2001}%
  \BibitemOpen
  \bibfield  {author} {\bibinfo {author} {\bibfnamefont {S.~M.}\ \bibnamefont
  {Bhattacharjee}}\ and\ \bibinfo {author} {\bibfnamefont {F.}~\bibnamefont
  {Seno}},\ }\href {\doibase 10.1088/0305-4470/34/33/302} {\bibfield  {journal}
  {\bibinfo  {journal} {Journal of Physics A: Mathematical and General}\
  }\textbf {\bibinfo {volume} {34}},\ \bibinfo {pages} {6375} (\bibinfo {year}
  {2001})}\BibitemShut {NoStop}%
\bibitem [{\citenamefont {Hata}\ \emph {et~al.}(2014)\citenamefont {Hata},
  \citenamefont {Nakao},\ and\ \citenamefont
  {Mikhailov}}]{hata_advection_2014}%
  \BibitemOpen
  \bibfield  {author} {\bibinfo {author} {\bibfnamefont {S.}~\bibnamefont
  {Hata}}, \bibinfo {author} {\bibfnamefont {H.}~\bibnamefont {Nakao}}, \ and\
  \bibinfo {author} {\bibfnamefont {A.~S.}\ \bibnamefont {Mikhailov}},\ }\href
  {\doibase 10.1103/PhysRevE.89.020801} {\bibfield  {journal} {\bibinfo
  {journal} {Physical Review E}\ }\textbf {\bibinfo {volume} {89}},\ \bibinfo
  {pages} {020801} (\bibinfo {year} {2014})}\BibitemShut {NoStop}%
\bibitem [{Note2()}]{Note2}%
  \BibitemOpen
  \bibinfo {note} {Generally, a network whose cost functions are monotonically
  increasing and involve a power law of degree no larger than $\alpha $ have $P
  \leq (\alpha +1)^{1+1/\alpha }/[(\alpha +1)^{1+1/\alpha } - \alpha ]$ \cite
  {roughgarden_price_2003}}\BibitemShut {NoStop}%
\bibitem [{\citenamefont {May}(1990)}]{may_traffic_1990}%
  \BibitemOpen
  \bibfield  {author} {\bibinfo {author} {\bibfnamefont {A.~D.}\ \bibnamefont
  {May}},\ }\href@noop {} {\emph {\bibinfo {title} {Traffic Flow
  Fundamentals}}}\ (\bibinfo  {publisher} {Prentice-Hall},\ \bibinfo {address}
  {Englewood Cliffs, {NJ}},\ \bibinfo {year} {1990})\BibitemShut {NoStop}%
\bibitem [{\citenamefont {Markoff}(2010)}]{markoff_google_2010}%
  \BibitemOpen
  \bibfield  {author} {\bibinfo {author} {\bibfnamefont {J.}~\bibnamefont
  {Markoff}},\ }\href
  {http://www.nytimes.com/2010/10/10/science/10google.html?pagewanted=all&_r=0}
  {\bibfield  {journal} {\bibinfo  {journal} {The New York Times}\ }\textbf
  {\bibinfo {volume} {10}},\ \bibinfo {pages} {A1} (\bibinfo {year}
  {2010})}\BibitemShut {NoStop}%
\end{thebibliography}%

\end{document}